\documentclass[journal=jpccck,manuscript=article]{achemso}

\newcommand{\hyd}{\textrm{H}_{2}}

\author{S{\"u}leyman Er}
\affiliation{Computational Materials Science, Faculty of Science and Technology and MESA+
Research Institute, University of Twente, P.O. Box 217, 7500 AE Enschede, The Netherlands}

\author{Gilles A. de Wijs}
\affiliation{Electronic Structure of Materials, Institute for
Molecules and Materials, Faculty of Science, Radboud University
Nijmegen, Heyendaalseweg 135, 6525 AJ Nijmegen, The Netherlands.}

\author{Geert Brocks}
\email{g.brocks@tnw.utwente.nl}
\affiliation{Computational Materials Science, Faculty of Science and Technology and MESA+
Research Institute, University of Twente, P.O. Box 217, 7500 AE Enschede, The Netherlands}

\title{DFT Study of Planar Boron Sheets: A New Template for Hydrogen Storage}

\begin{document}

\date{\today}

\begin{abstract}
We study the hydrogen storage properties of planar boron sheets and compare them to those of graphene. The binding of molecular hydrogen to the boron sheet ($0.05$ eV) is stronger than that to graphene. We find that dispersion of alkali metal (AM = Li, Na, and K) atoms onto the boron sheet markedly increases hydrogen binding energies and storage capacities. The unique structure of the boron sheet presents a template for creating a stable lattice of strongly bonded metal atoms with a large nearest neighbor distance. In contrast, AM atoms dispersed on graphene tend to cluster to form a bulk metal. In particular the boron-Li system is found to be a good candidate for hydrogen storage purposes. In the fully loaded case this compound can contain up to $10.7$ wt $\%$ molecular hydrogen with an average binding energy of $0.15$ eV/$\hyd$.
\end{abstract}
\maketitle

\section{Introduction}
Hydrogen is an abundant, clean, and renewable energy carrier \cite{coontz2004nss}. An important barrier preventing the large scale use of hydrogen is storing it densely and safely under moderate conditions \cite{schlapbach2001hsm}. Also, fast loading and unloading of the storage system is still a challenge. Storage of hydrogen in molecular form may be more beneficial than storage of atomic hydrogen in chemical hydrides. Complex metal hydrides, for instance, are often either too stable or too unstable \cite{vansetten2005prb,er2009ths} and require substantial doping to tune their stability \cite{vansetten2007dop,gremaud2007hoc}. Moreover, the formation and decomposition of chemical hydrides typically involve complicated solid-state chemical reactions, which hamper the kinetics of hydrogen loading and unloading. In contrast, materials that physisorb molecular hydrogen, such as graphene, carbon nanotubes and fullerenes, clathrates, zeolites, and metal organic frameworks (MOFs), are capable of achieving fast hydrogen kinetics. The binding energy of hydrogen to these host materials is however small, which leads to unfavorable operating conditions of very low temperatures or high hydrogen pressures \cite{bhatia2006oca}.

Recent studies show that metal doping enhances the strength of binding between hydrogen molecules and physisorption materials \cite{yildirim2005tdc,durgun2008fcb}. An ideal metal dopant should strongly bind to the host material, and it should bind hydrogen molecules effectively. For instance, for storage of molecular hydrogen at 30 bar under ambient temperature, and delivery at 1.5 bar, the optimum adsorption enthalpy needs to be 0.15 eV/H$_2$ \cite{bhatia2006oca}. Upon deposition, the geometrical positioning of metal atoms over the substrate is also important, since the number of $\hyd$ molecules per dopant site should be maximized. To prevent steric hindrance, this usually means that the dopant atoms need to be far apart. Therefore, the material should resist the clustering of metal atoms. At least the binding energy of a metal atom to the substrate should then be larger than the cohesive energy of the bulk metal. This is extremely difficult to attain with transition metal dopants, since the cohesive energies of bulk transition metals are high. Alkali metals have much lower cohesive energies and are therefore more suitable as dopants.

Recent studies addressed the alkali metal doping of physisorption materials such as graphene \cite{deng2004nad, ataca2008hch, er2009hsp}, fullerene \cite{sun2006fps, chandrakumar2008c60, sun2009apl}, carbon nanotubes \cite{chen1999hhu, liu1999hss, yang2000hsa, lee2002fps}, and MOFs \cite{blomqvist2007mof5, han2007idm, mavrandonakis2008wld, klontzas2008ihs}. Hydrogen molecules are trapped around the dispersed doping metal atoms through electrostatic and polarization interactions \cite{lochan2006}. In addition to increasing the binding strengths to hydrogen, alkali metal doping also increases the effective surface area available for absorption in some cases. Hydrogen adsorption then goes beyond a single monolayer, increasing the volumetric and gravimetric hydrogen densities accordingly.

In order to obtain a sufficiently high gravimetric density, not only the alkali metals need to be lightweight, but also the substrates need to be lightweight materials. In the past decade, carbon nanostructures of various dimensions have been studied extensively both theoretically and experimentally for various applications, including as substrates for hydrogen storage. The recent realization of the structurally most simple carbon network, graphene, has attracted a great deal of interest \cite{novoselov2004efe, geim2007gec}. The lightest element that can form extensive covalently bonded structures is in fact boron. Boron, having one electron less than carbon, forms icosahedral clusters,\cite{muetterties1967acb, werheit2000bor} and has long been thought not to form two-dimensional (2-D) structures similar to graphene.

\begin{figure}[!tb]
	\centering
		\includegraphics[width=8cm]{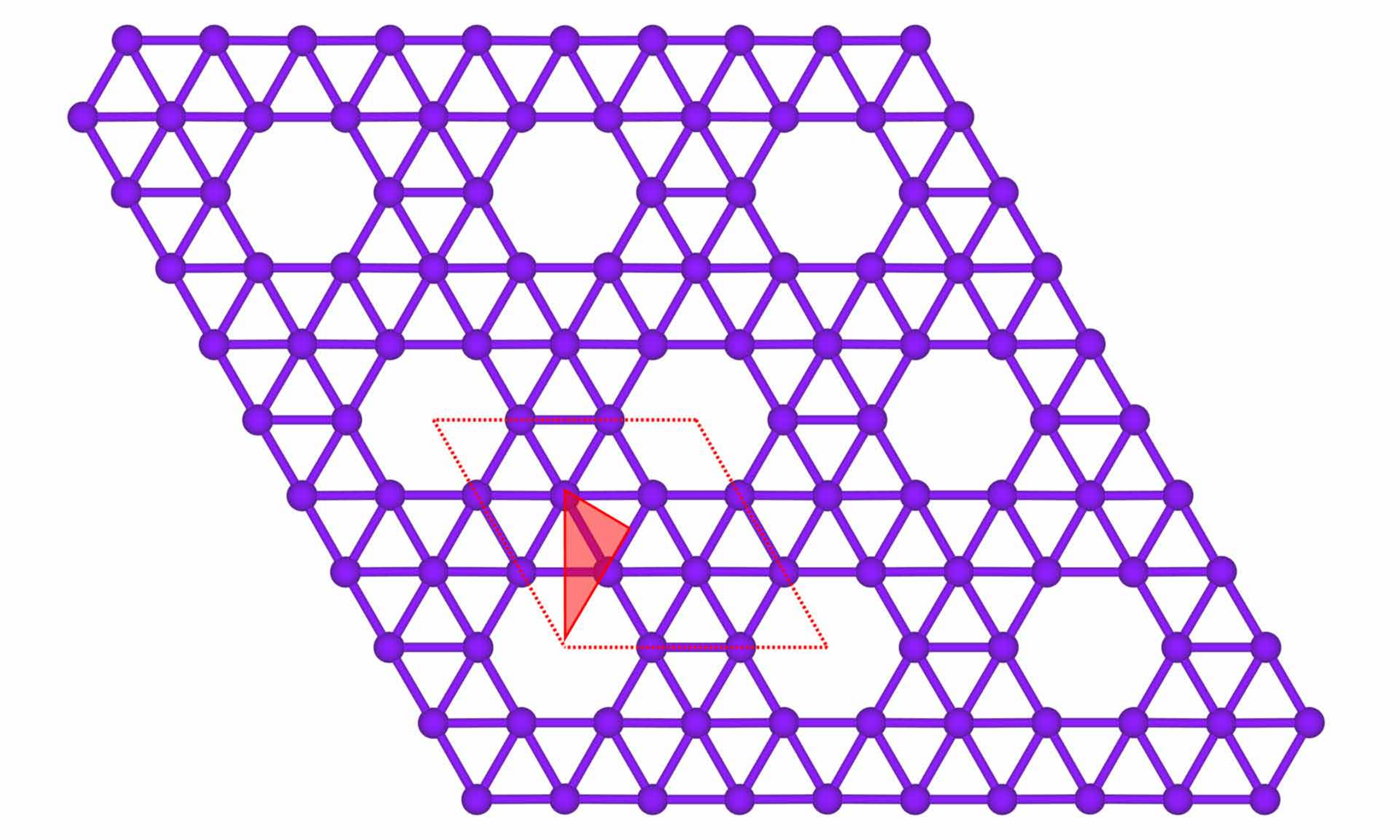}
		\caption{The most stable form of the boron sheet. The primitive cell is indicated by dotted red lines. It contains eight boron atoms, shown as purple spheres. The irreducible part of the Wigner$-$Seitz cell (WSC) is indicated by a triangle.}
	\label{fig:B8}
\end{figure}

However, it has very recently been shown that the stability of planar 2-D hexagonal boron sheets is enhanced markedly by inserting additional boron atoms in the centers of hexagons of the honeycomb structure; see \ref{fig:B8} \cite{tang2007npb, lau2007sae, lau2008tsn, yang2008aip}. These additional boron atoms modify the electronic structure of the boron lattice. 2-D hexagonal boron is an electron deficient system with part of the sp$^2$ bonding states unoccupied. The electrons provided by the additional boron atoms occupy these states, increasing the stability \cite{tang2007npb}. These new 2-D boron structures also provide insight into the unusual stability of other boron-only structures with different dimensions, such as fullerenes \cite{prasad2008sis, yan2008fbf, baruah2008vsa}, nanotubes \cite{singh2008ppb, yang2008aip, lau2008fps}, and nanoribbons \cite{ding2008esb}.

Here, we investigate the hydrogen storage properties of the boron sheet structure that is identified as the most stable form, the so-called $\alpha$-structure.\cite{tang2007npb} We first study the interaction of $\hyd$ molecules with this boron sheet. Next, the dispersion of alkali metal (AM = Li, Na, and K) atoms onto the boron sheet is considered. We study the binding and the mobility of these AM atoms on the boron sheet and the hydrogenation energies of these systems. All of the three AMs are found to bind strongly to the boron sheet. Moreover, the unique structure of the boron sheet allows for a stable lattice of AM atoms with a large nearest neighbor AM$-$AM distance, which consequently creates space for hydrogenation. The AM atoms accumulate a net positive charge, which not only stabilizes this structure but also strengthens the binding to hydrogen molecules. All three AMs lead to a higher hydrogen binding energy, but Li is by far the best doping element.

\section{Computational Details}
We perform first-principles calculations based on density functional theory (DFT), using the Vienna \textit{ab initio} simulation package (VASP) \cite{kresse1993aim, kresse1996eis}. The generalized gradient approximation (GGA) in the form of the PW91 functional is used to approximate exchange and correlation \cite{perdew1992ams}. Although dispersion (van der Waals) interactions are not captured by this functional, it should give a good description of chemical bonding and of electrostatic interactions, which are the most prominent interactions in alkali doped systems; see also ref  \cite{er2009hsp}. Binding energies of $\hyd$ molecules to similar systems obtained with the PW91 functional are within $\sim$20 meV/$\hyd$ of those obtained with the PBE functional \cite{er2009hsp}. Therefore, we only give the PW91 results in the following.

The basis set is constructed according to the projector augmented wave (PAW) method \cite{blochl1994paw, kresse1999upp}. The following electrons are treated as valence: H 1s, B 2s2p, Li 1s2s, Na 3s2p, and K 4s3p. A cutoff energy of $400$ eV is used for the plane-wave basis. Brillouin zones (BZs) of all structures are integrated with \textit{k}-point spacings of $\sim$0.01 \upshape{\AA}$^{-1}$ \cite{monkhorst1976spb}. For instance, the BZ corresponding to the unit cell shown in Figure \ref{fig:B8} is then integrated using a 12 $\times$ 12 regular grid. Gaussian smearing is used with a smearing width of 0.01 eV. Periodic images of the sheet along the surface normal are well separated by a distance of $\sim$20 \upshape{\AA}, so that the interactions between these images are negligible. Atomic and geometric relaxations are carried out by employing the conjugated-gradient (CG) algorithm. All atomic positions and cell parameters are relaxed. The relaxations are assumed to be complete when the total force remaining on each atom and the stress tensor components acting on the cells are less than $0.02$ eV/\upshape{\AA} and $2.5$ kbar, respectively. Total energies then are converged to within 1 meV/unit cell at least. The tetrahedron method is then used to calculate the total energies of the relaxed structures \cite{blochl1994itm}.

\section{Results and Discussion}
\subsection{Hydrogen Adsorption on Boron Sheets}

The most stable structure of a boron sheet is shown in \ref{fig:B8}. It is a
graphene-like structure, where two-thirds of the hexagon centers are occupied by additional boron atoms. The primitive cell of the sheet then contains eight boron atoms. It leads to a unique planar structure with a lattice of empty hexagons spaced at a distance of 5.06 \AA\ of one another.

\begin{figure}[!tb]
	\centering
		\includegraphics[width=8cm]{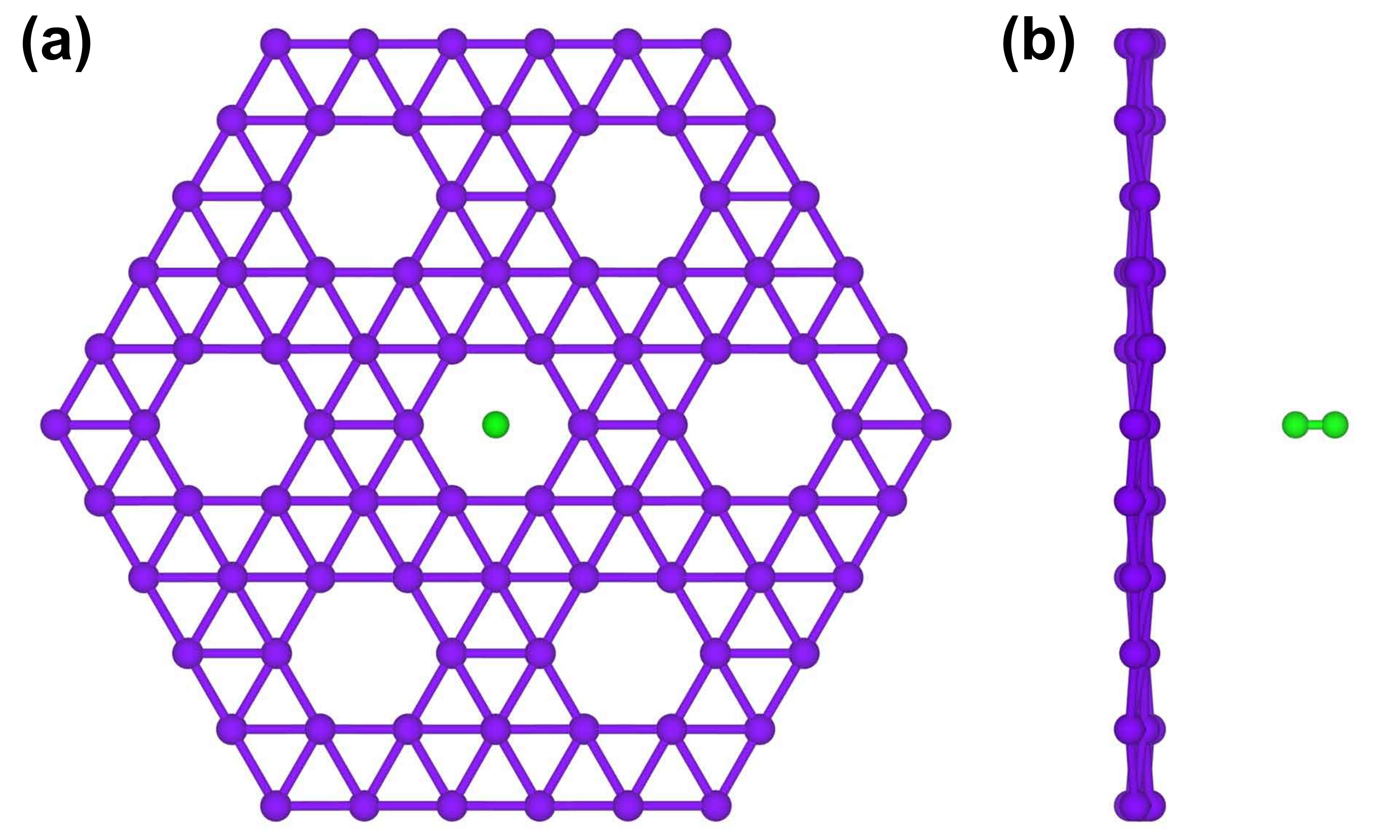}
		\caption{The most stable geometry of $\hyd$ adsorbed on a boron sheet shown in (a) top, and (b) side views. The molecule is positioned over the center of an empty boron hexagon with its axis orthogonal to the boron plane.}
	\label{fig:B32H2}
\end{figure}

To model the interaction of hydrogen molecules with the boron sheet, we use a 2 $\times$ 2 supercell. The distance between hydrogen molecules in neighboring cells is then $10.12$ \upshape{\AA}, which is sufficiently large to neglect the interaction between these molecules. We consider a number of possible positions of the $\hyd$ molecule on the boron sheet. Since the latter contains a lot of symmetry, it is only necessary to consider the irreducible part of the Wigner$-$Seitz cell (WSC), which is indicated in \ref{fig:B8}. We scan the edges of the irreducible WSC by placing a $\hyd$ molecule at several positions, which are separated by $0.8$ \upshape{\AA}, as measured from the H$-$H bond centers. At each of these points, we consider three different molecular orientations. In two of these orientations, the $\hyd$ molecular axis is parallel to the boron plane. The two orientations differ by an in-plane angle of $30$$^{\circ}$. In the third configuration, the $\hyd$ molecular axis is perpendicular to the boron plane. The H$-$H bond centers are placed at a distance of at least $2$ \upshape{\AA} from the boron sheet.

In total, we have constructed a set of $36$ distinct initial geometries. All of the structures in this set are then subjected to relaxation without any symmetry constraints. These calculations reveal that $\hyd$ molecules tend to occupy positions over the open boron hexagons. The minimum energy configuration is the one in which the molecule is over the center of an open hexagon with its axis perpendicular to the boron plane (\ref{fig:B32H2}). The distance between the boron plane and the center of the molecule is $3.27$ \upshape{\AA}, and the calculated value of the binding energy is $47$ meV. The energy differences between different orientations of the $\hyd$ molecule are generally small. A $\hyd$ molecule at this site with its axis parallel to the boron plane is only $\sim 5$ meV higher in energy. The potential energy surface obtained by varying the position of the $\hyd$ molecule over the plane has a corrugation of 61 meV.

Interestingly, the interaction of $\hyd$ with the boron sheet shows similarities to its interaction with graphene. We consider a graphene supercell similar to that of boron, which has of course the typical graphene structure without atoms in the centers of the hexagons. The calculated binding energy of $\hyd$ to graphene is then $25$ meV. Likewise, the geometry of the adsorbed $\hyd$ molecule is very similar to that on the boron sheet, and the distance to the graphene plane is $3.24$ \upshape{\AA}. Our results on the graphene$-$$\hyd$ system are in accordance with earlier studies \cite{heine2004hsp, henwood2007aii, park2007cbg}. We note that, although the binding of $\hyd$ to the boron sheet is slightly stronger than that to graphene, it is still too weak for a practical hydrogen storage application.

\subsection{Doping Boron Sheets with Alkali Metals}
\begin{figure}[!tb]
	\centering
		\includegraphics[width=8cm]{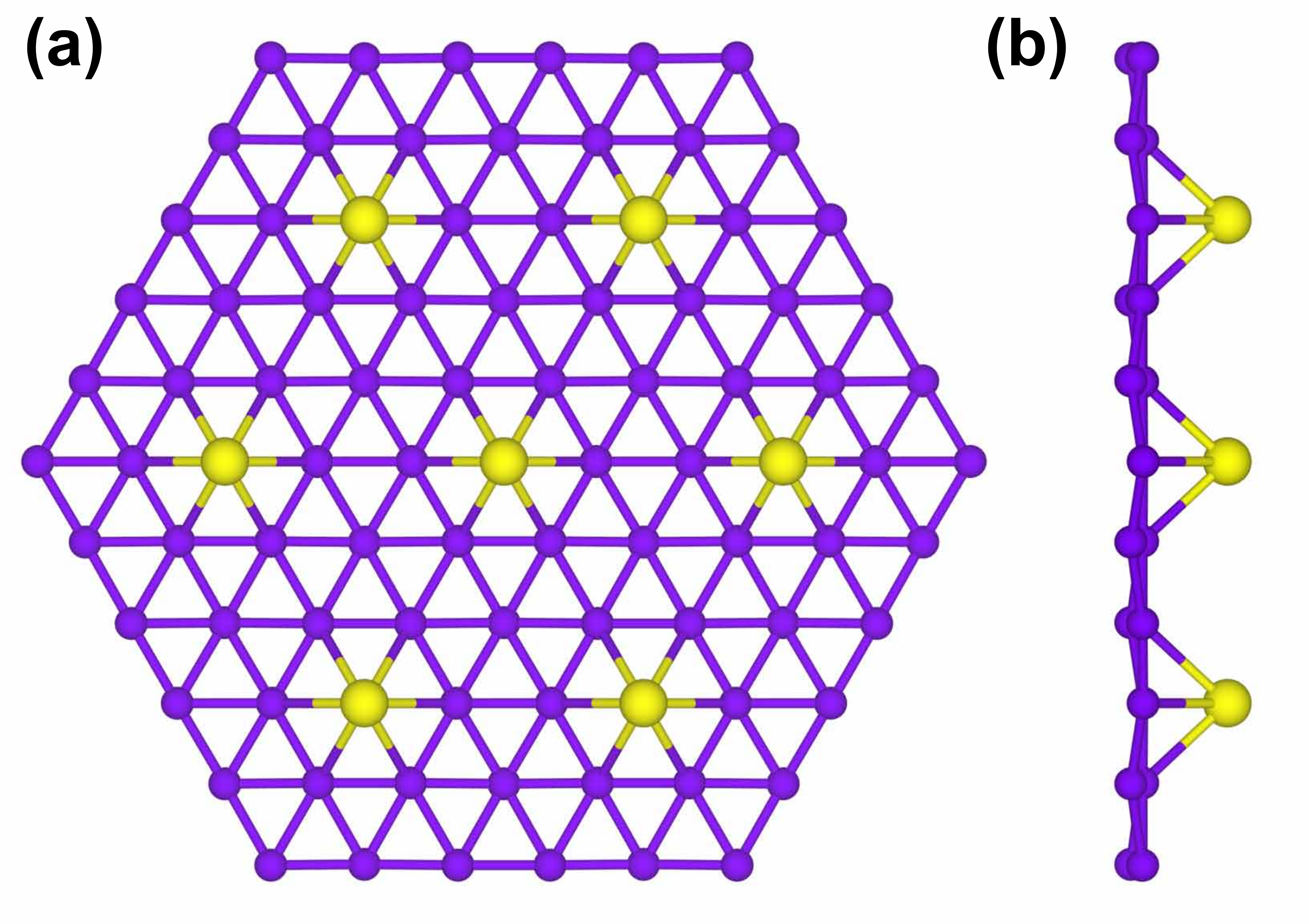}
		\caption{Optimized structure of a boron sheet doped with Li atoms (yellow spheres), shown from (a) the top and (b) the side. Boron sheets doped with other lightweight AMs have a similar geometry.}
	\label{fig:B8Li1}
\end{figure}

Next, we consider deposition of AM atoms onto the boron sheet, with AM = Li, Na, or K. To have a stable material at a reasonable temperature, it is essential that the metals strongly bind to the boron template. In addition, meaningful hydrogen gravimetric densities can be reached only if the AM atoms are separated far enough from each other so that each atom has sufficient space to bind hydrogen molecules. Placing a single AM atom in the unit cell shown in \ref{fig:B8} leads to a lattice of AM atoms with a cell parameter of $5.06$ \upshape{\AA}. We vary the position of the AM atom on the boron sheet and relax all structures, as discussed in the previous section. The optimal position of all three lightweight AMs turns out to be over the center of the empty boron hexagon, as in \ref{fig:B8Li1}. Upon relaxing, the unit cell parameters of the doped systems do not change noticeably. The optimized cell parameters are $5.08$, $5.07$, and $5.07$ \upshape{\AA} for the Li, Na, and K doped boron sheet, respectively.

The binding energies of the AM atoms to the boron sheet are calculated according to
\begin{equation}
E_{\rm b} = E_{\rm AMbulk} + E_{\rm boron} - E_{\rm boronAM},
\label{eq:eb}
\end{equation}
where $E_{\rm AMbulk}$ is the total energy per atom of the AM bulk metal, $E_{\rm boronAM}$ is the total energy per AM atom of the AM doped boron sheet, and $E_{\rm boron}$ is the corresponding total energy of the boron sheet. The results are shown in \ref{table:borAM1}. For comparison, the binding energies $E_{\rm at}$, calculated from eq \ref{eq:eb} by replacing $E_{\rm AMbulk}$ with the energy of an isolated AM atom, are also given in this table. The numbers show that all three systems are thermodynamically stable with respect to separation into bulk metal and a boron sheet.

The equilibrium distance between the AM atoms and the boron plane, listed in \ref{table:borAM1}, shows an increase going from Li to K, as might be expected. Electrostatic interactions are dominant in binding hydrogen molecules to the AM atoms \cite{henkelman2006far, tang2009jpcm}. The Bader charges decrease going from Li to K, which at first sight may seem surprising as the electronegativity of the AMs decreases along this series. However, comparing the nearest neighbor distances between the metal atoms on the boron surface with those in the bulk metals, one observes that the former are larger by 70, 40, and 10 \% for Li, Na, and K, respectively. This means that the Li atoms on the boron sheet are relatively isolated, but the K atoms are still in relatively close contact. In this configuration, a Li atom is bonded to the boron sheet and the binding has a large ionic contribution, whereas a K atom is bonded to the boron sheet and to other K atoms with a mix of ionic and metallic bonding. Clearly, the binding of a Na atom is intermediate between these two situations.

\begin{table}[!tb]
\centering
\caption{Calculated Binding Energies of Alkali Metals on a Boron Sheet (\ref{fig:B8Li1}) with Respect to Bulk Metal ($E_{\rm b}$), Isolated Atoms ($E_{\rm at}$), Equilibrium Distances to the Boron Plane ($d$), and Bader Charges on the Alkali Atoms ($Q$)}
\begin{tabular}{lcccc}
\hline
AM & $E_{\rm b}$(eV)& $E_{\rm at}$(eV)& $d$(\upshape{\AA})& $Q$ ($e$) \\
\hline
Li        & $0.27$ & $1.89$     & $1.54$   & $+0.85$        \\
Na        & $0.02$ & $1.11$     & $2.26$   & $+0.60$         \\
K         & $0.34$ & $1.21$     & $2.75$   & $+0.47$        \\
\hline
\end{tabular}
\label{table:borAM1}
\end{table}

In principle, one might increase the distance between the K atoms by decreasing their concentration on the boron sheet. However, K is already the heaviest element of the AM series considered here, and a reduction of their concentration increases the number of boron atoms per metal atom. This would seriously decrease the gravimetric hydrogen density, since the amount of hydrogen adsorbed depends on the metal atoms only. In conclusion, since Li is the lightest element, binds strongly to the boron sheet, and discharges almost completely upon deposition, it is the best candidate as a dopant for hydrogen storage purposes.

Since the Li atoms on the boron carry an effective positive charge, one may expect a net repulsion between these atoms. We consider the binding energy $E_{\rm b}$, eq \ref{eq:eb}, and Bader charge $Q$ as a function of the Li concentration. The results are given in \ref{table:concentration}. In the concentration we have considered so far, the Li$-$Li nearest neighbor distance is 5.08 \AA. Doubling the boron surface unit cell in each direction and occupying only one empty hexagon by a Li atom decreases the concentration by a factor of 4 and decreases the Li$-$Li nearest neighbor distance by a factor of 2. Occupying not only all empty, but also all filled boron hexagons increases the Li concentration by a factor of 3. The Li$-$Li nearest neighbor distance is then multiplied $1/\sqrt(3)$. The binding energy indeed decreases with increasing concentration, confirming the net repulsion between the Li atoms. There is a corresponding slight decrease of the Bader charge on the Li atoms. Note however that, at all concentrations studied, the system is stable against phase separation into a sheet carrying a lower concentration of Li and bulk Li. The Li concentration can therefore be controlled by the amount of Li dispersed on the boron sheet.

\begin{table}[!tb]
\centering
\caption{Calculated Binding Energies with Respect to Bulk Metal ($E_{\rm b}$) and Bader Charges ($Q$) of Li on a Boron Sheet at Various Concentrations, Represented by the Nearest Neighbor Metal$-$Metal Distance $d_{\rm nn}$.}
\begin{tabular}{ccc}
\hline
$d_{\rm nn}$(\upshape{\AA}) & $E_{\rm b}$(eV) & $Q$ ($e$) \\
\hline
10.16       & $0.45$ & $+0.88$        \\
5.08        & $0.27$ & $+0.85$        \\
2.93        & $0.16$ & $+0.75$        \\
\hline
\end{tabular}
\label{table:concentration}
\end{table}

It is interesting to compare the Li doping of the boron sheet to that of a similar system, namely, Li doping of graphene, which has been proposed as a suitable hydrogen storage system \cite{deng2004nad, ataca2008hch}. We use a similar unit cell as for the boron sheet, containing six carbon atoms and three hexagons, one of which contains a Li atom. After relaxation, the lattice parameter of the graphene$-$Li cell is $4.28$ \upshape{\AA}, whereas the boron-Li cell parameter is slightly larger ($5.08$ \upshape{\AA}). We find that the binding of Li atoms to graphene is considerably weaker than the binding to the boron sheet. On graphene, the binding energy $E_{\rm b} = -0.61$ eV. Obviously, a negative value of the binding energy indicates that this system is thermodynamically unstable with respect to bulk Li and graphene, in contrast to the boron-Li system, cf. \ref{table:borAM1}.

Moreover, a Bader analysis carried out on the graphene-Li system reveals that only $0.64e$ is removed from Li atoms after deposition, which is significantly smaller than the case of the boron$-$Li system. These results suggest that the boron sheet is a more promising candidate for lightweight AM doping than pure graphene.

\vspace{-8pt}
\subsection{Mobility of Alkali Metal Atoms}
\vspace{-4pt}

\begin{figure}[!tb]
	\centering
		\includegraphics[width=14cm]{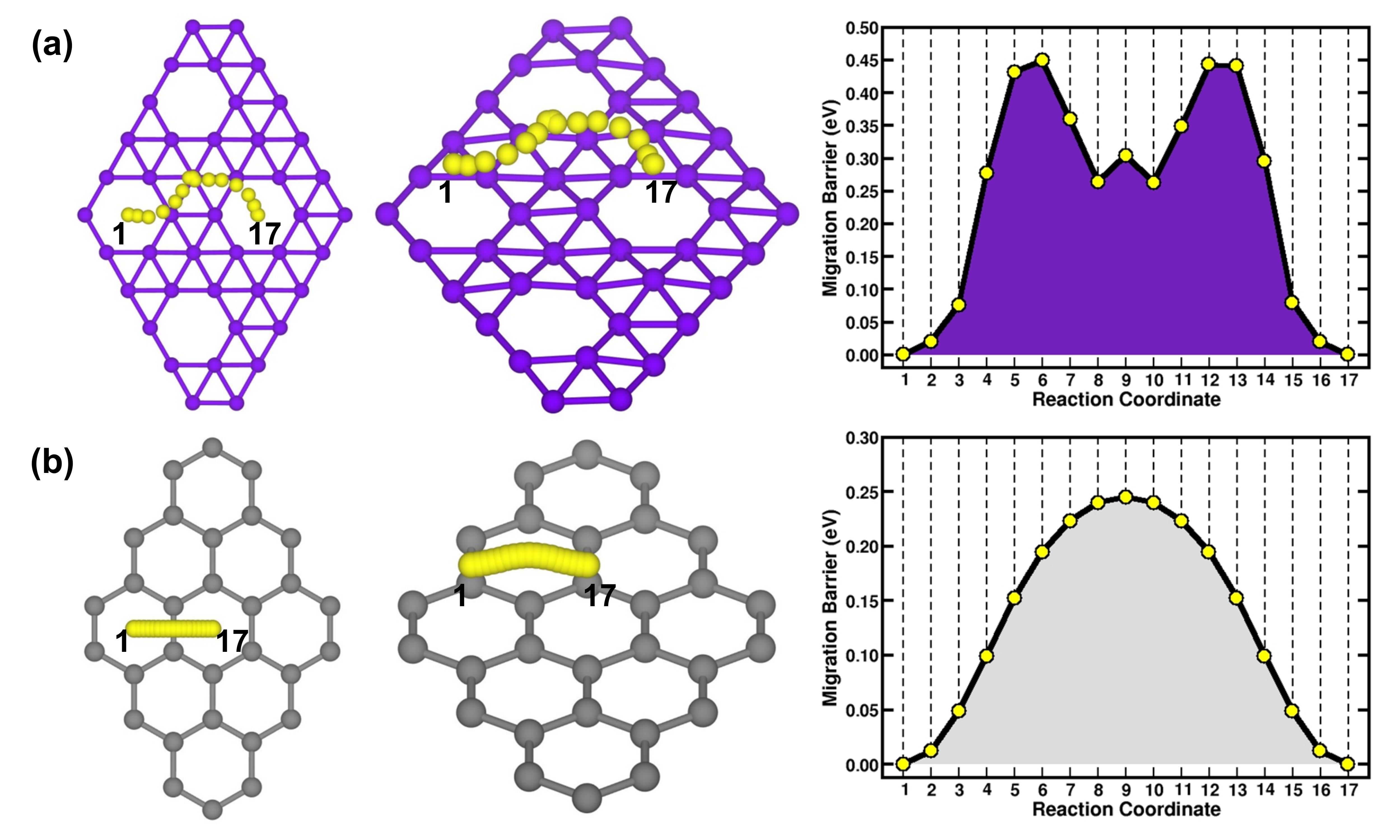}
		\caption{Diffusion of a Li atom over a (a) boron sheet and (b) graphene. The lowest energy diffusion paths are shown in top and perspective views. Calculated energy profiles along the paths are plotted on the right-hand side. The yellow dots in the plot correspond to the positions shown on the left-hand side. The diffusion barriers are 0.45 and 0.25 eV on the boron and graphene, respectively.}
	\label{fig:migration}
\vspace{-10pt}
\end{figure}

The stability of a material such as a doped boron or graphene sheet also depends on the mobility of the AM atoms on the sheet. Within the transition state theory model, the most important parameter determining the mobility is the energy barrier for diffusion. We determine the diffusion path of a single Li atom on the boron sheet using the nudged elastic band method \cite{jonsson1998neb, henkelman2000jcp, henkelman2008jcp}. The calculations are carried out using a 2 $\times$ 2 supercell with $32$ boron atoms, where one empty hexagon is occupied by a Li atom. A string of 17 images is chosen to model the diffusion path between two empty hexagons, as shown in \ref{fig:migration}a. Positions over the centers of the empty hexagons represent absolute energy minima, and positions over the boron atoms at the centers of the other hexagons represent secondary adsorption minima. An optimal path runs over B atoms, and along B$-$B bonds, as shown in \ref{fig:migration}a, with maxima over B$-$B bond centers. The calculated diffusion barrier is 0.45 eV. The lowest energy path for diffusion between two energy minima corresponding to empty hexagons proceeds via a secondary minimum over a filled hexagon, giving rise to a curved path. The simpler direct path between two empty hexagons involves an energy barrier of 0.55 eV and is therefore less likely.

We can compare this to diffusion of a Li atom over graphene. Positions over the centers of the carbon hexagons represent the absolute energy minima. A Li atom diffuses over a C$-$C bond following a simple path, with a maximum over a C$-$C bond center, as shown in \ref{fig:migration}b. The calculated diffusion barrier is $E_{\rm d}$ = 0.25 eV. Consequently, Li atoms on graphene are quite mobile. Using an Arrhenius expression for the jump rate $\nu = \nu_{0}$exp[$E_{\rm d}/kT$] with a typical attempt frequency of $\nu_{0} \approx 10^{12}$ Hz, one obtains $\nu \approx 10^{8}$ Hz at room temperature and $\nu \approx 1$ Hz at $T = 105$ K. Because of their negative binding energy with respect to bulk Li, as discussed in the previous section, there is a thermodynamic driving force for Li atoms to cluster, which would seriously hamper the use of doped graphene as a hydrogen storage material \cite{deng2004nad, ataca2008hch}. It is, however, possible to use graphene as a template for polylithiated molecules to create a stable storage material \cite{er2009hsp}.

The diffusion barrier, $E_{\rm d}$ = 0.45 eV, for Li atoms on the boron sheet is substantially larger than that on graphene. Using the same Arrhenius expression as above, a jump rate of $\nu \approx 1$ Hz is obtained at $T = 190$ K. At room temperature, individual Li atoms are still mobile ($\nu \approx 10^{4}$ Hz), but the formation bulk Li metal is thermodynamically unfavorable because of the high binding energy of the Li atoms on the boron sheet. The energy difference between the minimum energy sites of a Li atom (over the centers of empty boron hexagons) and the secondary minima is $0.3$ eV. It means that at room temperature only the minimum energy sites are occupied. Since the Li atoms carry a substantial effective charge, there is in addition no driving force for clustering of metal atoms on the sheet. The minimum energy sites on the boron sheet form a regular lattice with a spacing of 5.08 \AA\ between the lattice points, which forms the ideal template for creating a stable lattice of Li atoms.

\subsection{Hydrogen Adsorption on Alkali Doped Boron Sheets}

\begin{table}[!tb]
\centering
\caption{Calculated Binding Energies of Alkali Metals on a Boron Sheet Where Both Sides of the Sheet are Functionalized (\ref{fig:B8Li1}), with Respect to Bulk Metal ($E_{\rm b}$) and Isolated Atoms ($E_{\rm at}$), Equilibrium Distances to the Boron Plane ($d$), Bader Charges on the Alkali Atoms ($Q$), Hydrogen Binding Energies ($E_{\rm H_2}$), and Gravimetric Hydrogen densities of the B$_{8}$AM$_{2}$ Systems.}
\begin{tabular}{lcccccc}
\hline
AM & $E_{\rm b}$(eV) & $E_{\rm at}$(eV) & $d$(\upshape{\AA})& $Q$($e$)& $E_{\rm H_2}$(eV) & $\hyd$ (wt \%) \\
\hline
Li        & $0.18$ & $1.80$     & $1.62$   & $+0.82$        & $0.35-0.15$  & $10.75$      \\
Na        & $0.02$ & $1.11$     & $2.28$   & $+0.60$        & $0.13-0.07$  & $ 8.36$      \\
K         & $0.35$ & $1.22$     & $2.75$   & $+0.47$        & $0.06$       & $ 2.39$      \\
\hline
\end{tabular}
\label{table:borAM}
\end{table}

\begin{figure}[!tb]
	\centering
		\includegraphics[width=9.25cm]{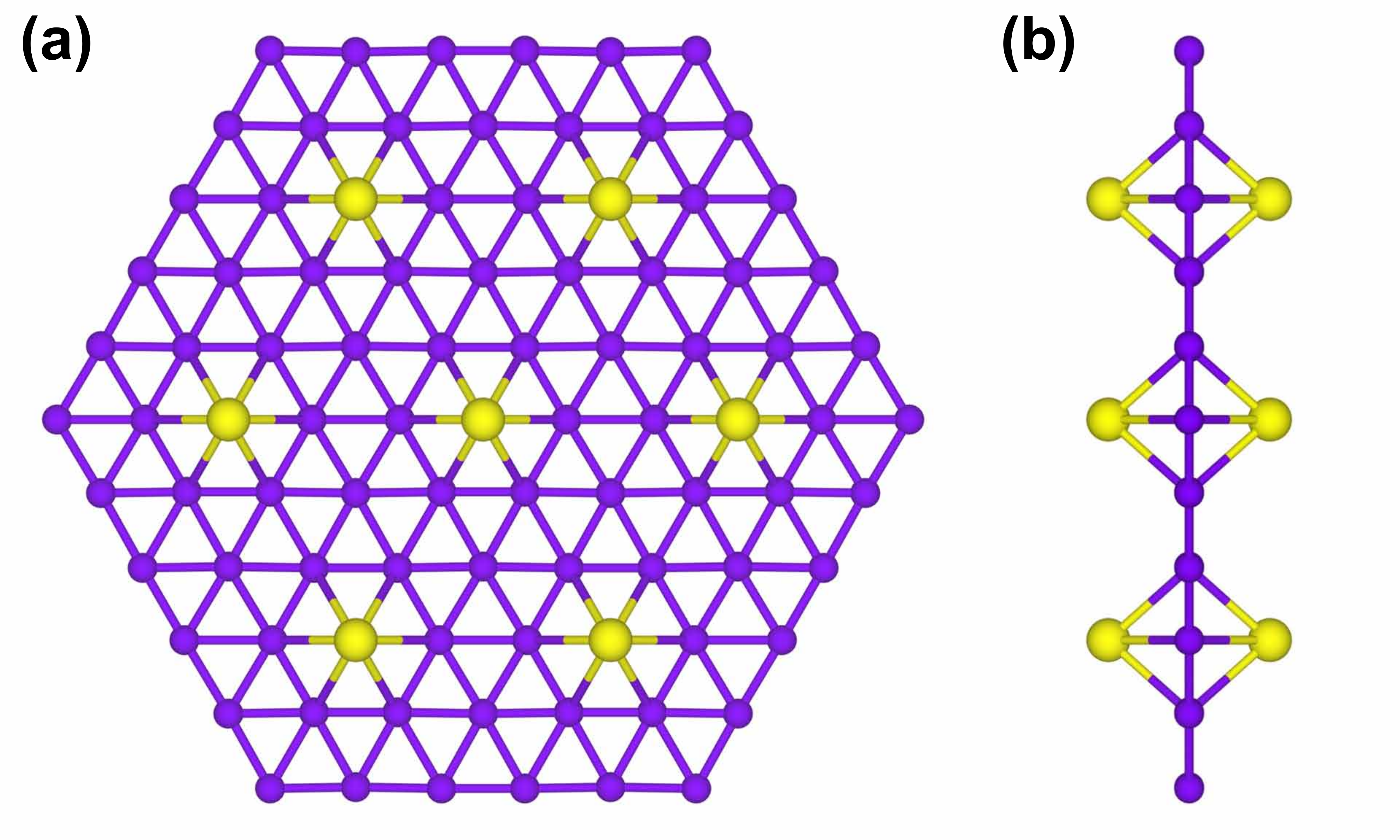}
		\caption{Optimized structure of a boron sheet doped with Li atoms on both sides, shown from (a) the top and (b) the side. Boron sheets doped with other lightweight AMs have a similar geometry.}
	\label{fig:B8Li2}
\end{figure}

Starting from the AM lattice on the boron sheet, as discussed in the previous sections, we find in fact that both faces of the boron lattice can be functionalized with AMs, as depicted in \ref{fig:B8Li2}. The chemical formula of the AM doped boron unit cells is then B$_{8}$AM$_{2}$ and the optimized lattice constants are $5.10$, $5.08$, and $5.08$ \upshape{\AA} for the AM = Li, Na, and K, respectively. The optimized structures have mirror symmetry with respect to the boron plane. The binding energies of the AM atoms, their distances to the boron plane, and their Bader charges are given in \ref{table:borAM}. A comparison to the data of \ref{table:borAM1} demonstrates that there is very little interaction between the AM atoms located above and below the boron sheet. Our general observations for AM doping remain valid; i.e., also for the doubly functionalized boron sheet, we have a stable lattice of effectively charged AM atoms.

Electrostatic and polarization interactions contribute most to the binding between the AM atoms and hydrogen molecules \cite{er2009hsp}. One may therefore expect that the effective charges on the metal atoms after deposition onto the boron sheet determines the strength of the interaction between the host material and the hydrogen molecules. These charges decrease along the series Li, Na, and K, as shown by \ref{table:borAM}. The binding energies of hydrogen molecules to the AM doped boron sheet follow this trend. The first two $\hyd$ molecules bind to B$_{8}$Li$_{2}$ with a binding energy of $0.35$ eV/$\hyd$ (one $\hyd$ binds to the top Li, and one to the bottom Li). The corresponding numbers for Na and K are significantly smaller, i.e., 0.13 and 0.06 eV/$\hyd$, respectively.

\begin{figure}[!tb]
	\centering
		\includegraphics[width=9.25cm]{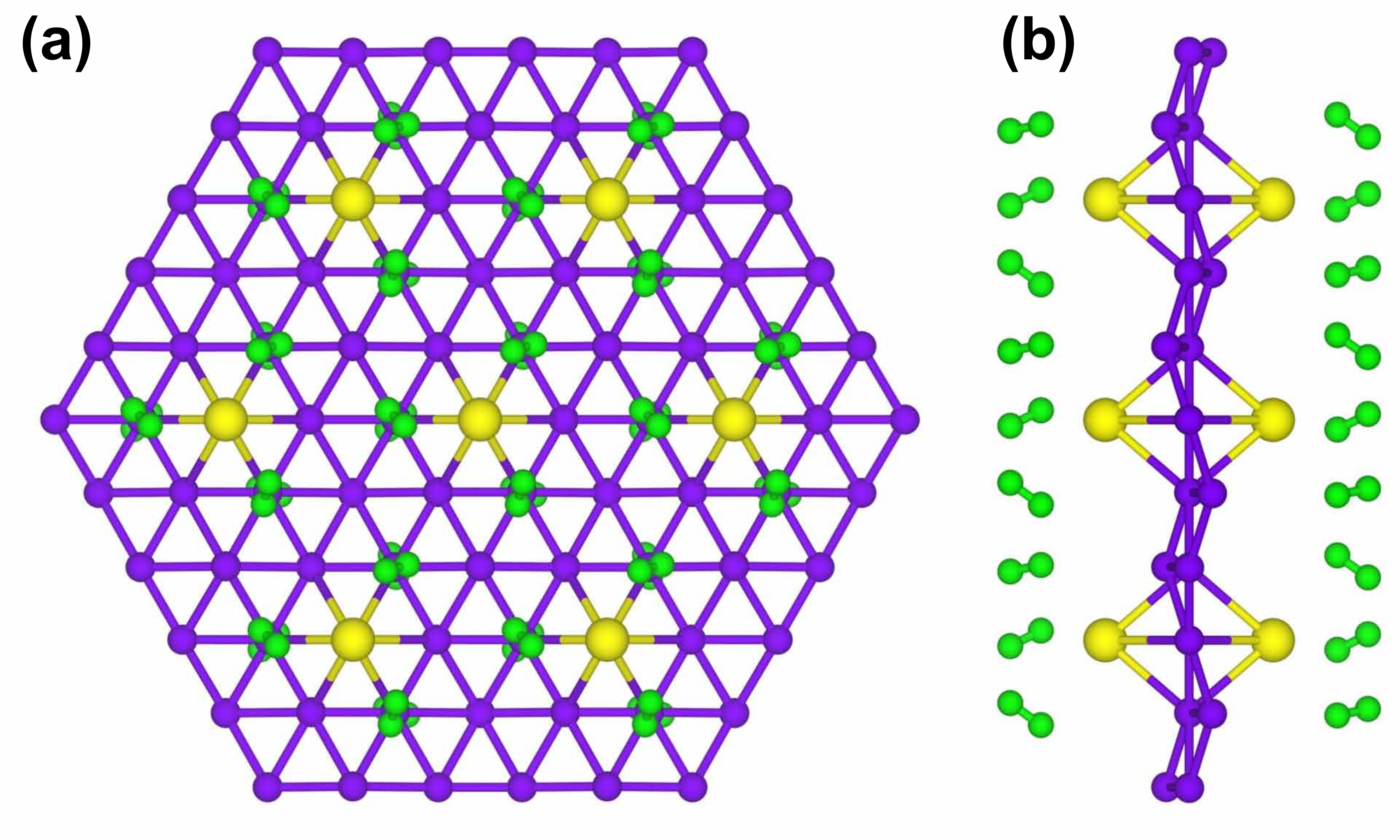}
		\caption{The optimized geometry of the Li doped boron sheet after full hydrogenation, shown from (a) the top and (b) the side. Three $\hyd$ molecules surround each Li atom.}
	\label{fig:B8Li2H12}
\end{figure}

\begin{figure}[!tb]
	\centering
		\includegraphics[width=9.25cm]{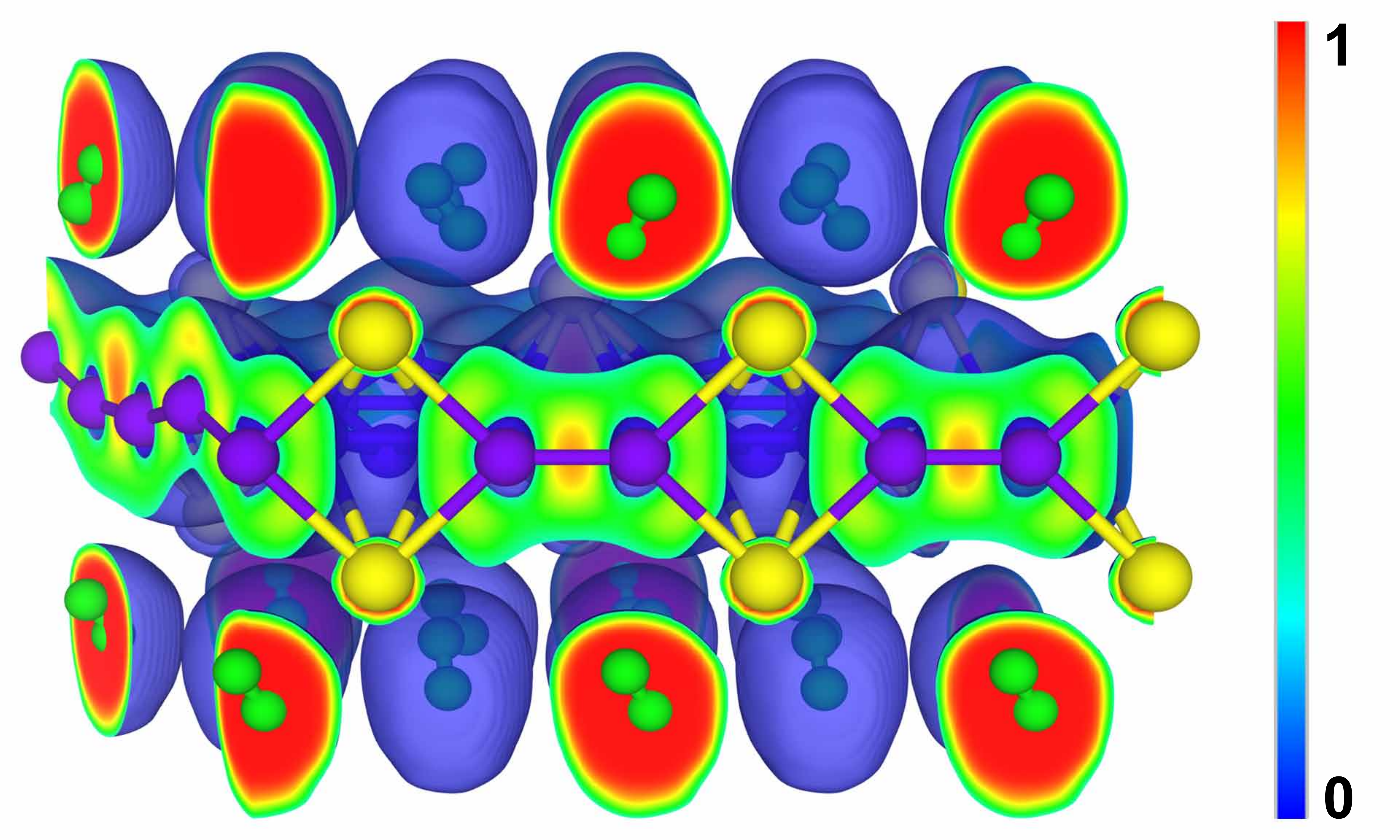}
		\caption{Electron localization function of the boron$-$Li system at its fully hydrogenated state. The electrons are localized at the $\hyd$ molecules and around the boron sheet, where the bonding is covalent. The positively charged Li ions are depleted from electrons.}
	\label{fig:elf}
\end{figure}

Each Li atom in this structure can capture up to three $\hyd$ molecules, but the average binding energy then drops to $0.15$ eV/$\hyd$. In the fully loaded case, the B$_{8}$Li$_{2}$ compound then contains $10.7$ wt $\%$ hydrogen, and the chemical formula of the unit cell is B$_{8}$Li$_{2}$(H$_{2}$)$_{6}$. The optimized geometry is shown in \ref{fig:B8Li2H12}. The centers of all the $\hyd$ molecules surrounding the Li atoms are located at an equivalent distance of $2.35$ \upshape{\AA} from the metal atoms. These results are in line with the interactions of $\hyd$ molecules with isolated AM cations \cite{lochan2006,chandrakumar2008c60}. The maximum number of $\hyd$ molecules that can be gathered around a metal center is determined by the Coulomb interactions that develop between Li$-$$\hyd$ and $\hyd$$-$$\hyd$, and by steric effects.

The B$_{8}$Na$_{2}$ system can also capture three $\hyd$ molecules per metal. The average binding energy for the $\hyd$ molecules then drops to $0.07$ eV/$\hyd$, which is still larger than the binding between a $\hyd$ molecule and a bare boron sheet. In contrast, the B$_{8}$K$_{2}$ system is capable of capturing only a single $\hyd$ molecule per K atom with a low binding energy of $0.06$ eV/$\hyd$. It is clear that, if compared to the Li doped system, the higher electron densities on the Na and K atoms result in weaker interactions with $\hyd$ molecules. The Li doped boron sheet is the best option for hydrogen storage.

It is interesting to note that the amount of charge on the Li atoms, as a consequence of their interaction with the boron lattice, stays almost constant during the consecutive addition of $\hyd$ molecules. The H$-$H bonds of the adsorbed $\hyd$ molecules are only elongated by $\sim 1.5$ \% as compared to an isolated $\hyd$ molecule, which illustrates that the interaction between $\hyd$ molecules and the Li atoms is mostly electrostatic and not chemical bonding.

The calculated electron localization function for the B$_{8}$Li$_{2}$(H$_{2}$)$_{6}$ system is shown in \ref{fig:elf}. The Li atoms are depleted, consistent with our Bader analysis. Hydrogens stay in molecular form and are surrounded by their localized electron clouds. There is no direct indication of an orbital interaction between the Li atoms and the hydrogens. Moreover, the pattern of the covalent bonding within the boron lattice is not significantly altered by the Li atoms or the $\hyd$ molecules.

\section{Conclusions}
We study the planar boron $\alpha$-sheet as a physisorption template for hydrogen storage. Direct physisorption of $\hyd$ on a boron sheet gives a binding energy of 0.05 eV. Alkali metal (AM) atoms can be dispersed on the boron sheet to increase the hydrogen binding energies. The boron sheet interacts strongly with AM atoms, and the doped systems are thermodynamically stable with respect to clustering of metal atoms. Moreover, the unique geometry of the boron plane provides a natural lattice for the metal atoms whose nearest neighbor distance can be tuned by varying their concentration. In particular, Li is found to be a promising doping element for hydrogen storage. The strong interaction between the boron sheet and the Li atoms results in an almost complete transfer of the Li valence electrons to the boron sheet. Electrostatic interactions between the well-exposed Li atoms and the $\hyd$ molecules then lead to an average binding energy of $0.15$ eV/$\hyd$. The system physisorbs up to a maximum of $10.7$ wt $\%$ hydrogen. In contrast, AM atoms dispersed on graphene are generally unstable with respect to the bulk metal, which suggests that AM doped nanostructures based upon boron are a much better starting point for hydrogen storage than their carbon based counterparts.

\vspace{-5pt}
\acknowledgement
\vspace{-10pt}
This work is part of the research programs of "Advanced Chemical Technologies for Sustainability (ACTS)" and the "Stichting voor Fundamenteel Onderzoek der Materie (FOM)". The use of supercomputer facilities was sponsored by the "Stichting Nationale Computerfaciliteiten (NCF)". These institutions are financially supported by "Nederlandse Organisatie voor Wetenschappelijk Onderzoek (NWO)".
\vspace{-20pt}
\bibliography{BORv09}

\ifx\mcitethebibliography\mciteundefinedmacro
\PackageError{achemso.bst}{mciteplus.sty has not been loaded}
{This bibstyle requires the use of the mciteplus package.}\fi
\begin{mcitethebibliography}{54}
\providecommand*{\natexlab}[1]{#1}
\mciteSetBstSublistMode{f}
\mciteSetBstMaxWidthForm{subitem}{(\alph{mcitesubitemcount})}
\mciteSetBstSublistLabelBeginEnd{\mcitemaxwidthsubitemform\space}
{\relax}{\relax}

\bibitem[coo()]{coontz2004nss}
 See the special issue \textit{Toward a Hydrogen Economy}, by R. Coontz and B.
  Hanson, Science \textbf{305}, 957 (2004)\relax
\mciteBstWouldAddEndPuncttrue
\mciteSetBstMidEndSepPunct{\mcitedefaultmidpunct}
{\mcitedefaultendpunct}{\mcitedefaultseppunct}\relax
\EndOfBibitem
\bibitem[Schlapbach and Z{\"u}ttel(2001)]{schlapbach2001hsm}
Schlapbach,~L.; Z{\"u}ttel,~A. \emph{Nature} \textbf{2001}, \emph{414},
  353--358\relax
\mciteBstWouldAddEndPuncttrue
\mciteSetBstMidEndSepPunct{\mcitedefaultmidpunct}
{\mcitedefaultendpunct}{\mcitedefaultseppunct}\relax
\EndOfBibitem
\bibitem[{van Setten} et~al.(2005){van Setten}, {de Wijs}, Popa, and
  Brocks]{vansetten2005prb}
{van Setten},~M.~J.; {de Wijs},~G.~A.; Popa,~V.~A.; Brocks,~G. \emph{Phys. Rev.
  B} \textbf{2005}, \emph{72}, 073107\relax
\mciteBstWouldAddEndPuncttrue
\mciteSetBstMidEndSepPunct{\mcitedefaultmidpunct}
{\mcitedefaultendpunct}{\mcitedefaultseppunct}\relax
\EndOfBibitem
\bibitem[Er et~al.(2009)Er, Tiwari, {de Wijs}, and Brocks]{er2009ths}
Er,~S.; Tiwari,~D.; {de Wijs},~G.~A.; Brocks,~G. \emph{Phys. Rev. B}
  \textbf{2009}, \emph{79}, 024105\relax
\mciteBstWouldAddEndPuncttrue
\mciteSetBstMidEndSepPunct{\mcitedefaultmidpunct}
{\mcitedefaultendpunct}{\mcitedefaultseppunct}\relax
\EndOfBibitem
\bibitem[{van Setten} et~al.(2007){van Setten}, {de Wijs}, and
  Brocks]{vansetten2007dop}
{van Setten},~M.~J.; {de Wijs},~G.~A.; Brocks,~G. \emph{Phys. Rev. B}
  \textbf{2007}, \emph{76}, 075125\relax
\mciteBstWouldAddEndPuncttrue
\mciteSetBstMidEndSepPunct{\mcitedefaultmidpunct}
{\mcitedefaultendpunct}{\mcitedefaultseppunct}\relax
\EndOfBibitem
\bibitem[Gremaud et~al.(2007)Gremaud, Broedersz, Borsa, Borgschulte, Mauron,
  Schreuders, Rector, Dam, and Griessen]{gremaud2007hoc}
Gremaud,~R.; Broedersz,~C.~P.; Borsa,~D.~M.; Borgschulte,~A.; Mauron,~P.;
  Schreuders,~H.; Rector,~J.~H.; Dam,~B.; Griessen,~R. \emph{Adv. Mater.}
  \textbf{2007}, \emph{19}, 2813\relax
\mciteBstWouldAddEndPuncttrue
\mciteSetBstMidEndSepPunct{\mcitedefaultmidpunct}
{\mcitedefaultendpunct}{\mcitedefaultseppunct}\relax
\EndOfBibitem
\bibitem[Bhatia and Myers(2006)]{bhatia2006oca}
Bhatia,~S.~K.; Myers,~A.~L. \emph{Langmuir} \textbf{2006}, \emph{22},
  1688--1700\relax
\mciteBstWouldAddEndPuncttrue
\mciteSetBstMidEndSepPunct{\mcitedefaultmidpunct}
{\mcitedefaultendpunct}{\mcitedefaultseppunct}\relax
\EndOfBibitem
\bibitem[Yildirim and Ciraci(2005)]{yildirim2005tdc}
Yildirim,~T.; Ciraci,~S. \emph{Phys. Rev. Lett.} \textbf{2005}, \emph{94},
  175501\relax
\mciteBstWouldAddEndPuncttrue
\mciteSetBstMidEndSepPunct{\mcitedefaultmidpunct}
{\mcitedefaultendpunct}{\mcitedefaultseppunct}\relax
\EndOfBibitem
\bibitem[Durgun et~al.(2008)Durgun, Ciraci, and Yildirim]{durgun2008fcb}
Durgun,~E.; Ciraci,~S.; Yildirim,~T. \emph{Phys. Rev. B} \textbf{2008},
  \emph{77}, 85405\relax
\mciteBstWouldAddEndPuncttrue
\mciteSetBstMidEndSepPunct{\mcitedefaultmidpunct}
{\mcitedefaultendpunct}{\mcitedefaultseppunct}\relax
\EndOfBibitem
\bibitem[Deng et~al.(2004)Deng, Xu, and Goddard]{deng2004nad}
Deng,~W.~Q.; Xu,~X.; Goddard,~W.~A. \emph{Phys. Rev. Lett.} \textbf{2004},
  \emph{92}, 166103\relax
\mciteBstWouldAddEndPuncttrue
\mciteSetBstMidEndSepPunct{\mcitedefaultmidpunct}
{\mcitedefaultendpunct}{\mcitedefaultseppunct}\relax
\EndOfBibitem
\bibitem[Ataca et~al.(2008)Ataca, Akt{\"u}rk, Ciraci, and
  Ustunel]{ataca2008hch}
Ataca,~C.; Akt{\"u}rk,~E.; Ciraci,~S.; Ustunel,~H. \emph{Appl. Phys. Lett.}
  \textbf{2008}, \emph{93}, 043123\relax
\mciteBstWouldAddEndPuncttrue
\mciteSetBstMidEndSepPunct{\mcitedefaultmidpunct}
{\mcitedefaultendpunct}{\mcitedefaultseppunct}\relax
\EndOfBibitem
\bibitem[Er et~al.(2009)Er, {de Wijs}, and Brocks]{er2009hsp}
Er,~S.; {de Wijs},~G.~A.; Brocks,~G. \emph{J. Phys. Chem. C} \textbf{2009},
  \emph{113}, 8997--9002\relax
\mciteBstWouldAddEndPuncttrue
\mciteSetBstMidEndSepPunct{\mcitedefaultmidpunct}
{\mcitedefaultendpunct}{\mcitedefaultseppunct}\relax
\EndOfBibitem
\bibitem[Sun et~al.(2006)Sun, Jena, Wang, and Marquez]{sun2006fps}
Sun,~Q.; Jena,~P.; Wang,~Q.; Marquez,~M. \emph{J. Am. Chem. Soc.}
  \textbf{2006}, \emph{128}, 9741--9745\relax
\mciteBstWouldAddEndPuncttrue
\mciteSetBstMidEndSepPunct{\mcitedefaultmidpunct}
{\mcitedefaultendpunct}{\mcitedefaultseppunct}\relax
\EndOfBibitem
\bibitem[Chandrakumar and Ghosh(2008)]{chandrakumar2008c60}
Chandrakumar,~K. R.~S.; Ghosh,~S.~K. \emph{Nano Lett.} \textbf{2008}, \emph{8},
  13--19\relax
\mciteBstWouldAddEndPuncttrue
\mciteSetBstMidEndSepPunct{\mcitedefaultmidpunct}
{\mcitedefaultendpunct}{\mcitedefaultseppunct}\relax
\EndOfBibitem
\bibitem[Sun et~al.(2009)Sun, Wang, and Jena]{sun2009apl}
Sun,~Q.; Wang,~Q.; Jena,~P. \emph{Appl. Phys. Lett.} \textbf{2009}, \emph{94},
  013111\relax
\mciteBstWouldAddEndPuncttrue
\mciteSetBstMidEndSepPunct{\mcitedefaultmidpunct}
{\mcitedefaultendpunct}{\mcitedefaultseppunct}\relax
\EndOfBibitem
\bibitem[Chen et~al.(1999)Chen, Wu, Lin, and Tan]{chen1999hhu}
Chen,~P.; Wu,~X.; Lin,~J.; Tan,~K.~L. \emph{Science} \textbf{1999}, \emph{285},
  91--93\relax
\mciteBstWouldAddEndPuncttrue
\mciteSetBstMidEndSepPunct{\mcitedefaultmidpunct}
{\mcitedefaultendpunct}{\mcitedefaultseppunct}\relax
\EndOfBibitem
\bibitem[Liu et~al.(1999)Liu, Fan, Liu, Cong, Cheng, and
  Dresselhaus]{liu1999hss}
Liu,~C.; Fan,~Y.~Y.; Liu,~M.; Cong,~H.~T.; Cheng,~H.~M.; Dresselhaus,~M.~S.
  \emph{Science} \textbf{1999}, \emph{286}, 1127--1129\relax
\mciteBstWouldAddEndPuncttrue
\mciteSetBstMidEndSepPunct{\mcitedefaultmidpunct}
{\mcitedefaultendpunct}{\mcitedefaultseppunct}\relax
\EndOfBibitem
\bibitem[Yang(2000)]{yang2000hsa}
Yang,~R. \emph{Carbon} \textbf{2000}, \emph{38}, 623--626\relax
\mciteBstWouldAddEndPuncttrue
\mciteSetBstMidEndSepPunct{\mcitedefaultmidpunct}
{\mcitedefaultendpunct}{\mcitedefaultseppunct}\relax
\EndOfBibitem
\bibitem[Lee et~al.(2002)Lee, Kim, Jin, and Chang]{lee2002fps}
Lee,~E.; Kim,~Y.; Jin,~Y.; Chang,~K. \emph{Phys. Rev. B} \textbf{2002},
  \emph{66}, 73415--73415\relax
\mciteBstWouldAddEndPuncttrue
\mciteSetBstMidEndSepPunct{\mcitedefaultmidpunct}
{\mcitedefaultendpunct}{\mcitedefaultseppunct}\relax
\EndOfBibitem
\bibitem[Blomqvist et~al.(2007)Blomqvist, Araujo, Srepusharawoot, and
  Ahuja]{blomqvist2007mof5}
Blomqvist,~A.; Araujo,~C.~M.; Srepusharawoot,~P.; Ahuja,~R. \emph{Proc. Natl.
  Acad. Sci. U.S.A.} \textbf{2007}, \emph{104}, 20173--20176\relax
\mciteBstWouldAddEndPuncttrue
\mciteSetBstMidEndSepPunct{\mcitedefaultmidpunct}
{\mcitedefaultendpunct}{\mcitedefaultseppunct}\relax
\EndOfBibitem
\bibitem[Han et~al.(2007)Han, Deng, and Goddard]{han2007idm}
Han,~S.~S.; Deng,~W.~Q.; Goddard,~W.~A. \emph{Angew. Chem., Int. Ed.}
  \textbf{2007}, \emph{46}, 6289--92\relax
\mciteBstWouldAddEndPuncttrue
\mciteSetBstMidEndSepPunct{\mcitedefaultmidpunct}
{\mcitedefaultendpunct}{\mcitedefaultseppunct}\relax
\EndOfBibitem
\bibitem[Mavrandonakis et~al.(2008)Mavrandonakis, Tylianakis, Stubos, and
  Froudakis]{mavrandonakis2008wld}
Mavrandonakis,~A.; Tylianakis,~E.; Stubos,~A.~K.; Froudakis,~G.~E. \emph{J.
  Phys. Chem. C} \textbf{2008}, \emph{112}, 7290--7294\relax
\mciteBstWouldAddEndPuncttrue
\mciteSetBstMidEndSepPunct{\mcitedefaultmidpunct}
{\mcitedefaultendpunct}{\mcitedefaultseppunct}\relax
\EndOfBibitem
\bibitem[Klontzas et~al.(2008)Klontzas, Mavrandonakis, Tylianakis, and
  Froudakis]{klontzas2008ihs}
Klontzas,~E.; Mavrandonakis,~A.; Tylianakis,~E.; Froudakis,~G. \emph{Nano
  Lett.} \textbf{2008}, \emph{8}, 1572--1576\relax
\mciteBstWouldAddEndPuncttrue
\mciteSetBstMidEndSepPunct{\mcitedefaultmidpunct}
{\mcitedefaultendpunct}{\mcitedefaultseppunct}\relax
\EndOfBibitem
\bibitem[Lochan and Head-Gordon(2006)]{lochan2006}
Lochan,~R.~C.; Head-Gordon,~M. \emph{Phys. Chem. Chem. Phys.} \textbf{2006},
  \emph{8}, 1357--1370\relax
\mciteBstWouldAddEndPuncttrue
\mciteSetBstMidEndSepPunct{\mcitedefaultmidpunct}
{\mcitedefaultendpunct}{\mcitedefaultseppunct}\relax
\EndOfBibitem
\bibitem[Novoselov et~al.(2004)Novoselov, Geim, Morozov, Jiang, Zhang, Dubonos,
  Grigorieva, and Firsov]{novoselov2004efe}
Novoselov,~K.~S.; Geim,~A.~K.; Morozov,~S.~V.; Jiang,~D.; Zhang,~Y.;
  Dubonos,~S.~V.; Grigorieva,~I.~V.; Firsov,~A.~A. \emph{Science}
  \textbf{2004}, \emph{306}, 666--669\relax
\mciteBstWouldAddEndPuncttrue
\mciteSetBstMidEndSepPunct{\mcitedefaultmidpunct}
{\mcitedefaultendpunct}{\mcitedefaultseppunct}\relax
\EndOfBibitem
\bibitem[Geim and MacDonald(2007)]{geim2007gec}
Geim,~A.; MacDonald,~A. \emph{Phys. Today} \textbf{2007}, \emph{60},
  35--41\relax
\mciteBstWouldAddEndPuncttrue
\mciteSetBstMidEndSepPunct{\mcitedefaultmidpunct}
{\mcitedefaultendpunct}{\mcitedefaultseppunct}\relax
\EndOfBibitem
\bibitem[Muetterties(1967)]{muetterties1967acb}
Muetterties,~E.~L. \emph{{The chemistry of boron and its compounds}};
\newblock New York: John Wiley \& Sons, Inc., 1967\relax
\mciteBstWouldAddEndPuncttrue
\mciteSetBstMidEndSepPunct{\mcitedefaultmidpunct}
{\mcitedefaultendpunct}{\mcitedefaultseppunct}\relax
\EndOfBibitem
\bibitem[Werheit(2000)]{werheit2000bor}
Werheit,~H. \emph{{Landolt-B{\"o}rnstein: Numerical Data and Functional
  Relationships in Science and Technology - New Series}};
\newblock Berlin: Springer, 2000;
\newblock pp 1--491\relax
\mciteBstWouldAddEndPuncttrue
\mciteSetBstMidEndSepPunct{\mcitedefaultmidpunct}
{\mcitedefaultendpunct}{\mcitedefaultseppunct}\relax
\EndOfBibitem
\bibitem[Tang and Ismail-Beigi(2007)]{tang2007npb}
Tang,~H.; Ismail-Beigi,~S. \emph{Phys. Rev. Lett.} \textbf{2007}, \emph{99},
  115501\relax
\mciteBstWouldAddEndPuncttrue
\mciteSetBstMidEndSepPunct{\mcitedefaultmidpunct}
{\mcitedefaultendpunct}{\mcitedefaultseppunct}\relax
\EndOfBibitem
\bibitem[Lau and Pandey(2007)]{lau2007sae}
Lau,~K.~C.; Pandey,~R. \emph{J. Phys. Chem. C} \textbf{2007}, \emph{111},
  2906--2912\relax
\mciteBstWouldAddEndPuncttrue
\mciteSetBstMidEndSepPunct{\mcitedefaultmidpunct}
{\mcitedefaultendpunct}{\mcitedefaultseppunct}\relax
\EndOfBibitem
\bibitem[Lau and Pandey(2008)]{lau2008tsn}
Lau,~K.~C.; Pandey,~R. \emph{J. Phys. Chem. B} \textbf{2008}, \emph{112},
  10217--10220\relax
\mciteBstWouldAddEndPuncttrue
\mciteSetBstMidEndSepPunct{\mcitedefaultmidpunct}
{\mcitedefaultendpunct}{\mcitedefaultseppunct}\relax
\EndOfBibitem
\bibitem[Yang et~al.(2008)Yang, Ding, and Ni]{yang2008aip}
Yang,~X.; Ding,~Y.; Ni,~J. \emph{Phys. Rev. B} \textbf{2008}, \emph{77},
  041402\relax
\mciteBstWouldAddEndPuncttrue
\mciteSetBstMidEndSepPunct{\mcitedefaultmidpunct}
{\mcitedefaultendpunct}{\mcitedefaultseppunct}\relax
\EndOfBibitem
\bibitem[Prasad and Jemmis(2008)]{prasad2008sis}
Prasad,~D.~L.; Jemmis,~E.~D. \emph{Phys. Rev. Lett.} \textbf{2008}, \emph{100},
  165504--165504\relax
\mciteBstWouldAddEndPuncttrue
\mciteSetBstMidEndSepPunct{\mcitedefaultmidpunct}
{\mcitedefaultendpunct}{\mcitedefaultseppunct}\relax
\EndOfBibitem
\bibitem[Yan et~al.(2008)Yan, Sheng, Zheng, Zhang, and Su]{yan2008fbf}
Yan,~Q.~B.; Sheng,~X.~L.; Zheng,~Q.~R.; Zhang,~L.~Z.; Su,~G. \emph{Phys. Rev.
  B} \textbf{2008}, \emph{78}, 20\relax
\mciteBstWouldAddEndPuncttrue
\mciteSetBstMidEndSepPunct{\mcitedefaultmidpunct}
{\mcitedefaultendpunct}{\mcitedefaultseppunct}\relax
\EndOfBibitem
\bibitem[Baruah et~al.(2008)Baruah, Pederson, and Zope]{baruah2008vsa}
Baruah,~T.; Pederson,~M.~R.; Zope,~R.~R. \emph{Phys. Rev. B} \textbf{2008},
  \emph{78}, 045408\relax
\mciteBstWouldAddEndPuncttrue
\mciteSetBstMidEndSepPunct{\mcitedefaultmidpunct}
{\mcitedefaultendpunct}{\mcitedefaultseppunct}\relax
\EndOfBibitem
\bibitem[Singh et~al.(2008)Singh, Sadrzadeh, and Yakobson]{singh2008ppb}
Singh,~A.~K.; Sadrzadeh,~A.; Yakobson,~B.~I. \emph{Nano Lett.} \textbf{2008},
  \emph{8}, 1314--1317\relax
\mciteBstWouldAddEndPuncttrue
\mciteSetBstMidEndSepPunct{\mcitedefaultmidpunct}
{\mcitedefaultendpunct}{\mcitedefaultseppunct}\relax
\EndOfBibitem
\bibitem[Lau et~al.(2008)Lau, Orlando, and Pandey]{lau2008fps}
Lau,~K.~C.; Orlando,~R.; Pandey,~R. \emph{J. Phys.: Condens. Matter}
  \textbf{2008}, \emph{20}, 125202\relax
\mciteBstWouldAddEndPuncttrue
\mciteSetBstMidEndSepPunct{\mcitedefaultmidpunct}
{\mcitedefaultendpunct}{\mcitedefaultseppunct}\relax
\EndOfBibitem
\bibitem[Ding et~al.(2008)Ding, Yang, and Ni]{ding2008esb}
Ding,~Y.; Yang,~X.; Ni,~J. \emph{Appl. Phys. Lett.} \textbf{2008}, \emph{93},
  043107\relax
\mciteBstWouldAddEndPuncttrue
\mciteSetBstMidEndSepPunct{\mcitedefaultmidpunct}
{\mcitedefaultendpunct}{\mcitedefaultseppunct}\relax
\EndOfBibitem
\bibitem[Kresse and Hafner(1993)]{kresse1993aim}
Kresse,~G.; Hafner,~J. \emph{Phys. Rev. B} \textbf{1993}, \emph{47},
  558--561\relax
\mciteBstWouldAddEndPuncttrue
\mciteSetBstMidEndSepPunct{\mcitedefaultmidpunct}
{\mcitedefaultendpunct}{\mcitedefaultseppunct}\relax
\EndOfBibitem
\bibitem[Kresse and Furthm{\"u}ller(1996)]{kresse1996eis}
Kresse,~G.; Furthm{\"u}ller,~J. \emph{Phys. Rev. B} \textbf{1996}, \emph{54},
  11169--11186\relax
\mciteBstWouldAddEndPuncttrue
\mciteSetBstMidEndSepPunct{\mcitedefaultmidpunct}
{\mcitedefaultendpunct}{\mcitedefaultseppunct}\relax
\EndOfBibitem
\bibitem[Perdew et~al.(1992)Perdew, Chevary, Vosko, Jackson, Pederson, Singh,
  and Fiolhais]{perdew1992ams}
Perdew,~J.~P.; Chevary,~J.~A.; Vosko,~S.~H.; Jackson,~K.~A.; Pederson,~M.~R.;
  Singh,~D.~J.; Fiolhais,~C. \emph{Phys. Rev. B} \textbf{1992}, \emph{46},
  6671--6687\relax
\mciteBstWouldAddEndPuncttrue
\mciteSetBstMidEndSepPunct{\mcitedefaultmidpunct}
{\mcitedefaultendpunct}{\mcitedefaultseppunct}\relax
\EndOfBibitem
\bibitem[Bl{\"o}chl(1994)]{blochl1994paw}
Bl{\"o}chl,~P.~E. \emph{Phys. Rev. B} \textbf{1994}, \emph{50},
  17953--17979\relax
\mciteBstWouldAddEndPuncttrue
\mciteSetBstMidEndSepPunct{\mcitedefaultmidpunct}
{\mcitedefaultendpunct}{\mcitedefaultseppunct}\relax
\EndOfBibitem
\bibitem[Kresse and Joubert(1999)]{kresse1999upp}
Kresse,~G.; Joubert,~D. \emph{Phys. Rev. B} \textbf{1999}, \emph{59},
  1758--1775\relax
\mciteBstWouldAddEndPuncttrue
\mciteSetBstMidEndSepPunct{\mcitedefaultmidpunct}
{\mcitedefaultendpunct}{\mcitedefaultseppunct}\relax
\EndOfBibitem
\bibitem[Monkhorst and Pack(1976)]{monkhorst1976spb}
Monkhorst,~H.~J.; Pack,~J.~D. \emph{Phys. Rev. B} \textbf{1976}, \emph{13},
  5188--5192\relax
\mciteBstWouldAddEndPuncttrue
\mciteSetBstMidEndSepPunct{\mcitedefaultmidpunct}
{\mcitedefaultendpunct}{\mcitedefaultseppunct}\relax
\EndOfBibitem
\bibitem[Bl{\"o}chl et~al.(1994)Bl{\"o}chl, Jepsen, and
  Andersen]{blochl1994itm}
Bl{\"o}chl,~P.~E.; Jepsen,~O.; Andersen,~O.~K. \emph{Phys. Rev. B}
  \textbf{1994}, \emph{49}, 16223--16233\relax
\mciteBstWouldAddEndPuncttrue
\mciteSetBstMidEndSepPunct{\mcitedefaultmidpunct}
{\mcitedefaultendpunct}{\mcitedefaultseppunct}\relax
\EndOfBibitem
\bibitem[Heine et~al.(2004)Heine, Zhechkov, and Seifert]{heine2004hsp}
Heine,~T.; Zhechkov,~L.; Seifert,~G. \emph{Phys. Chem. Chem. Phys.}
  \textbf{2004}, \emph{6}, 980--984\relax
\mciteBstWouldAddEndPuncttrue
\mciteSetBstMidEndSepPunct{\mcitedefaultmidpunct}
{\mcitedefaultendpunct}{\mcitedefaultseppunct}\relax
\EndOfBibitem
\bibitem[Henwood and Carey(2007)]{henwood2007aii}
Henwood,~D.; Carey,~J.~D. \emph{Phys. Rev. B} \textbf{2007}, \emph{75},
  245413\relax
\mciteBstWouldAddEndPuncttrue
\mciteSetBstMidEndSepPunct{\mcitedefaultmidpunct}
{\mcitedefaultendpunct}{\mcitedefaultseppunct}\relax
\EndOfBibitem
\bibitem[Park et~al.(2007)Park, Hong, Kim, and Jhi]{park2007cbg}
Park,~N.; Hong,~S.; Kim,~G.; Jhi,~S.~H. \emph{J. Am. Chem. Soc.} \textbf{2007},
  \emph{129}, 8999--9003\relax
\mciteBstWouldAddEndPuncttrue
\mciteSetBstMidEndSepPunct{\mcitedefaultmidpunct}
{\mcitedefaultendpunct}{\mcitedefaultseppunct}\relax
\EndOfBibitem
\bibitem[Henkelman et~al.(2006)Henkelman, Arnaldsson, and
  J{\'o}nsson]{henkelman2006far}
Henkelman,~G.; Arnaldsson,~A.; J{\'o}nsson,~H. \emph{Comput. Mater. Sci.}
  \textbf{2006}, \emph{36}, 354--360\relax
\mciteBstWouldAddEndPuncttrue
\mciteSetBstMidEndSepPunct{\mcitedefaultmidpunct}
{\mcitedefaultendpunct}{\mcitedefaultseppunct}\relax
\EndOfBibitem
\bibitem[Tang et~al.(2009)Tang, Sanville, and Henkelman]{tang2009jpcm}
Tang,~W.; Sanville,~E.; Henkelman,~E. \emph{J. Phys.: Condens. Matter}
  \textbf{2009}, \emph{21}, 084204\relax
\mciteBstWouldAddEndPuncttrue
\mciteSetBstMidEndSepPunct{\mcitedefaultmidpunct}
{\mcitedefaultendpunct}{\mcitedefaultseppunct}\relax
\EndOfBibitem
\bibitem[J\'{o}nsson et~al.(1998)J\'{o}nsson, Mills, and
  Jacobsen]{jonsson1998neb}
J\'{o}nsson,~H.; Mills,~G.; Jacobsen,~K.~W. \emph{{Nudged Elastic Band Method
  for Finding Minimum Energy Paths of Transitions, in Classical and Quantum
  Dynamics in Condensed Phase Simulations}};
\newblock Singapore: World Scientific, 1998;
\newblock p 385\relax
\mciteBstWouldAddEndPuncttrue
\mciteSetBstMidEndSepPunct{\mcitedefaultmidpunct}
{\mcitedefaultendpunct}{\mcitedefaultseppunct}\relax
\EndOfBibitem
\bibitem[Henkelman and J{\'o}nsson(2000)]{henkelman2000jcp}
Henkelman,~G.; J{\'o}nsson,~H. \emph{J. Chem. Phys.} \textbf{2000}, \emph{113},
  9978\relax
\mciteBstWouldAddEndPuncttrue
\mciteSetBstMidEndSepPunct{\mcitedefaultmidpunct}
{\mcitedefaultendpunct}{\mcitedefaultseppunct}\relax
\EndOfBibitem
\bibitem[Sheppard et~al.(2008)Sheppard, Terrell, and
  Henkelman]{henkelman2008jcp}
Sheppard,~D.; Terrell,~R.; Henkelman,~G. \emph{J. Chem. Phys.} \textbf{2008},
  \emph{128}, 134106\relax
\mciteBstWouldAddEndPuncttrue
\mciteSetBstMidEndSepPunct{\mcitedefaultmidpunct}
{\mcitedefaultendpunct}{\mcitedefaultseppunct}\relax
\EndOfBibitem
\end{mcitethebibliography}

\end{document}